\documentclass[letter]{aa}
\usepackage{natbib}
\bibpunct{(}{)}{;}{a}{}{,} % to follow the A&A style
\usepackage{graphicx}
\usepackage{amsmath}
\usepackage{multirow}
\usepackage{subcaption}
\usepackage{txfonts}
\newcommand{\Msol}{M_\odot}

\begin{document} 

\title{
Parallax in microlensing toward the Magellanic Clouds: \\
Effect on detection efficiency and detectability}
\author{
T.~Blaineau\inst{1},
M.~Moniez\inst{1}
}
\institute{
Laboratoire de physique des 2 infinis Ir\`ene Joliot-Curie,
CNRS Universit\'e Paris-Saclay,
B\^at. 100, Facult\'e des sciences, F-91405 Orsay Cedex, France
}

\offprints{M. Moniez,\\ \email{ moniez@lal.in2p3.fr}}
%{\it see also our WWW server at  URL:} \\
%{\tt http://www.lal.in2p3.fr/recherche/eros}}

\date{Received 24/03/2020, accepted 10/04/2020}
%
%%%%%%%%%%%%%%%%%%%%%%%%%%%%%%%%%%%%%%%%%%%%%%%%%%
%                                                %
%    BEGINNING OF TEXT                           %
%                                                %
%%%%%%%%%%%%%%%%%%%%%%%%%%%%%%%%%%%%%%%%%%%%%%%%%%
%\maketitle
%\tableofcontents

\abstract
{}{
We study the effect of the parallax on the search for microlensing events on very long timescales towards the Magellanic Clouds caused by dark massive compact objects within the past MACHO and EROS, the on-going MOA and OGLE, and the future LSST surveys.
We quantify what neglecting this effect means in the classical event selection process and also quantify the parallax detectability without the help of follow-up observations.
\\
}
{
We defined the distance between true events affected by parallax and the closest events without parallax.
This distance was used to estimate the probability of missing the preselection of events because of parallax for any survey characterised by its time sampling and photometric performance.
We also defined another distance to quantify the detectability of the parallax effect in order to trigger complementary observations.
\\
}
{
We find that the preselection of year-long timescale events is marginally affected by parallax for all surveys if the criteria are reasonably tight.
We also show that the parallax should be detectable in most of the events found by the LSST survey without follow-up observations.
\\
}
{}


\titlerunning{Effect of parallax on microlensing detection}
\authorrunning{Blaineau, Moniez}

\maketitle

\section{Introduction}
The recently observed gravitational waves \citep{Abbott_2016b, Abbott_2016a} might be
emitted by possible candidates for the Galactic halo dark matter \citep{Bird_2016}.
The quest for direct evidence of intermediate-mass black holes has resulted in a revival of the long-timescale microlensing searches (more than a few years).

Several teams have operated systematic microlensing survey programs to search for hidden
massive
compact objects after the publication of Paczy\'nski \citep{Paczynski86}: the Exp\'erience de Recherche d'Objets Sombres (EROS) by \citet{Aubourg1993}, the survey MAssive Compact Halo Objects (MACHO) by \citet{Alcock1993},
the Optical Gravitational Lensing Experiment (OGLE) by \citet{OGLE1993}, and the survey called Microlensing Observations in Astrophysics (MOA) by \citet{Sako_2007}.
The global result is that objects lighter than $10 M_{\odot}$ contribute for a negligible fraction of the
Galactic spherical halo mass \citep{Tisserand_2007, Wyrzykowski_2011}.
Searches are now ongoing to explore the halo beyond this limit by searching for events with longer timescales
either by combining the databases (the Moa-Eros-Macho-Ogle (MEMO) project, \citet{Mirhosseini_2018}), by extending the surveys (OGLE), or by optimising
the strategy of the future survey with the Large Synoptic Survey Telescope (LSST) \citep{collaboration2009lsst}.
\\
In this letter, we examine one specificity of the long-timescale events, which is the distortions
expected from the orbital motion of Earth around the Sun (so-called parallax effect in this paper),
which can become significant for events longer than a few months.
In Sect. \ref{section:formalism} we compare a microlensing event as seen from the Sun and from Earth.
In Sect. \ref{section:simulation} we describe a procedure for simulating a representative sample of microlensing events that take parallax into account as expected from the distributions of Galactic hidden massive compact objects.
In Sect. \ref{section:detection} we propose a quantitative characterisation of the distortion induced
by the parallax with respect to the simple rectilinear motion events. We deduce the maximum effect on
the microlensing first-level filtering, tuned by considering only non-parallax events.
In Sect. \ref{section:potential} we define a proxy of the significance of parallax that can be used for any microlensing survey.
We evaluate the parallax detectability from
simple representations of surveys based on the time sampling and the photometric resolution
in two realistic cases: the MEMO and the LSST projects (Sect. \ref{section:discussion}).

In this paper, we focus on the microlensing detection towards the Large Magellanic Cloud (LMC), and the Small Magellanic Cloud (SMC),
where the dark matter signal from compact objects is expected mainly from a distribution model between a spherical halo and a thick disc.

\section{Microlensing effect and the parallax}
\label{section:formalism}
Microlensing occurs when a massive compact object passes close enough to the line of sight of a background source and temporarily magnifies its brightness.
A review of the microlensing formalism can be found in \citet{Schneider_2006} and \citet{Rahvar_2015}.
When a single point-like lens of mass $M$ located at distance $D_L$ deflects the
light from a point source located at distance $D_S$, the magnification $A(t)$
of the source luminosity is given by \citet{Paczynski86}
\begin{equation}
A(t)=\frac{u(t)^2+2}{u(t)\sqrt{u(t)^2+4}}\ ,
\label{Amplification}
\end{equation}
where $u(t)$ is the distance of the lensing object to the undeflected line of sight, divided by
the Einstein radius $r_{\mathrm{E}}$,
\begin{equation}
r_{\mathrm{E}}\! =\!\! \sqrt{\frac{4GM}{c^2}D_S x(1\! -\! x)}\!
\simeq\! 4.5\mathrm{AU}\!\left[\frac{M}{\Msol}\right]^{\frac{1}{2}}\!
\left[\frac{D_S}{10 kpc}\right]^{\frac{1}{2}}\!\!
\frac{\left[x(1\! -\! x)\right]^{\frac{1}{2}}}{0.5}\!. %\nonumber
\end{equation}
Here $G$ is the Newtonian gravitational constant, and $x = D_L/D_S$.
If the lens moves at a constant relative transverse velocity $v_T$ and $u(t)$ reaches its minimum
value $u_0$ (impact parameter) to the undeflected line of sight
at time $t_0$, then $u(t)=\sqrt{u_0^2+(t-t_0)^2/t_{\mathrm{E}}^2}$,
where $t_{\mathrm{E}}=r_{\mathrm{E}} /v_T$ is the lensing timescale,
\begin{equation}
t_{\mathrm{E}} \sim
79\ \mathrm{days} \times %\\
\left[\frac{v_T}{100\, km/s}\right]^{-1}
\left[\frac{M}{\Msol}\right]^{\frac{1}{2}}
\left[\frac{D_S}{10\, kpc}\right]^{\frac{1}{2}}
\frac{[x(1-x)]^{\frac{1}{2}}}{0.5}. %\nonumber
\label{eq.tE}
\end{equation}
$v_T$ combines the velocities of the source, lens, and observer. 
When the microlensing is observed from the Sun, $v_T$ is a constant, but when it is observed from the Earth, $v_T$ includes a rotating component that is responsible for the so-called parallax effect.

\subsection{Microlensing events observed from the Sun}
The so-called simple (or standard) microlensing effect (point-like source and lens
with rectilinear motions) has the following characteristic
features:
Because the event results from a very rare alignment, it should occur only once for the source (as well as the deflector) regardless of the monitoring duration;
the magnification $A(t)$ does not depend on the colour
and is a function of time
depending only on ($u_0, t_0, t_{\mathrm{E}}$),
with a symmetric shape.
The prior distribution of the events' impact parameters $u_0$ is uniform;
and all stars at the same given distance have the same probability of being lensed.
This simple microlensing description can be altered by
many different effects, for example by parallax when observed from Earth or from a satellite instead of the Sun \citep{Alcock_1995, Han_1995, Novati_2016}, multiple lens and source systems \citep{Mao_1995},
extended sources \citep{Yoo_2004}, etc.

\subsection{Microlensing events observed from Earth}
\begin{figure}[htbp]
\begin{center}
    \includegraphics[width=9cm]{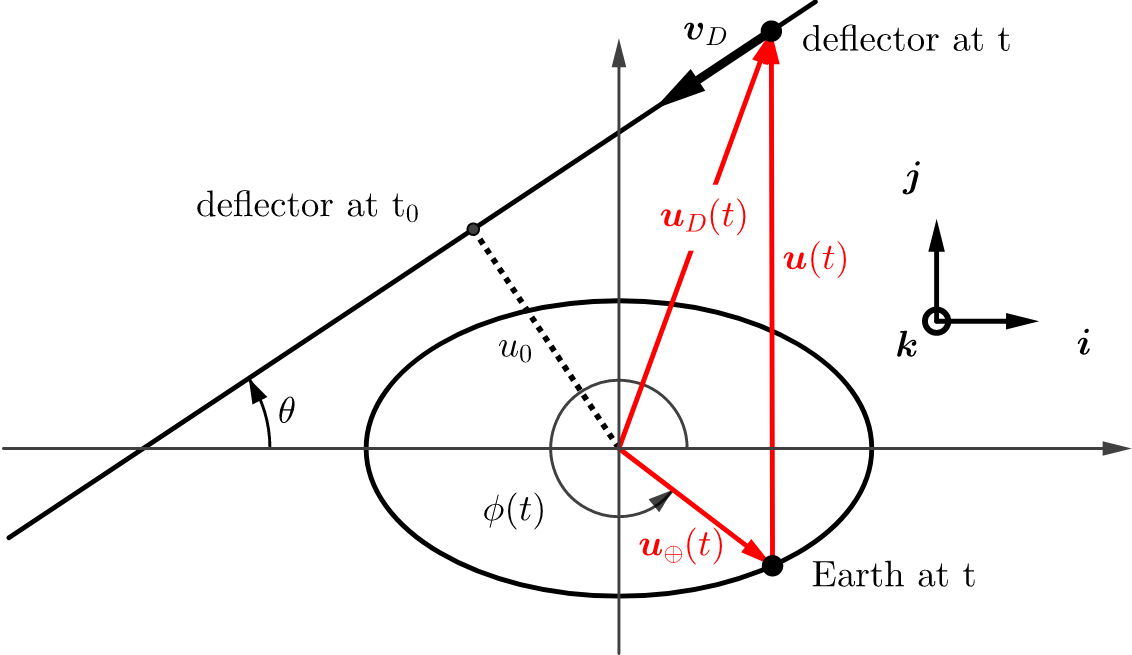}
    \caption[]{\it Parallax in microlensing. Projection from the source (here in the northern ecliptic hemisphere) on the plane ({\bf i, j}), perpendicular to the Sun-source direction {\bf k} containing the Sun.\\
    {\bf i} is along the intersection of the ecliptic and transverse planes. Its orientation is such that the perpendicular projection of {\bf j} on the ecliptic plane is opposite to the source. 
    }
\label{parallax}
\end{center}
\end{figure}

The orbital motion of Earth induces a curved (non-rectilinear) relative motion of the deflector with respect to the line of sight \citep{Gould_1992}.
Figure \ref{parallax} shows the configuration of a microlensing event as seen from the source, projected in the transverse plane containing the Sun. 
To take the parallax into account, two additional parameters are needed. These are
\begin{itemize}
    \item$\pi_E = \dfrac{r_\oplus}{r_E}(1-x)$: the ratio between the orbital radius of Earth and the projected Einstein radius and
    \item $\theta\in\left[0, 2\pi \right]$: the angle between {\bf i} and the transverse speed vector of the deflector (see Fig. \ref{parallax}).
\end{itemize}
Because the invariance of a microlensing event with respect to the moving direction of the lens is broken, we account for this by allowing $u_0$ to take negative value when the kinetic momentum ${\bf u}_D(t)\wedge{\bf v}_D$ is opposite to {\bf k}.
The reference time $t_0$ is now the time of minimum approach of the deflector to the line of sight from the Sun (not from Earth).
The positions of the projections of Earth and the deflector in the ({\bf i, j, k}) frame are expressed in units of projected Einstein radius $r_E/(1-x)$ as
\begin{equation}
\pmb{u}_\oplus(t)=
\begin{bmatrix}
    \pi_E\cos{\phi(t)} \\
    \pi_E \sin{\phi(t)}\sin{\beta_\star}\\
    0
\end{bmatrix}, 
\pmb{u}_D(t) = 
\begin{bmatrix}
    -u_0 \sin{\theta} - \frac{t-t_0}{t_E}\cos{\theta}\\
    u_0 \cos{\theta} - \frac{t-t_0}{t_E}\sin{\theta}\\
    0
\end{bmatrix},
\end{equation}
where $\phi(t)$ is defined in Fig. \ref{parallax}, and $\beta_\star$ is the ecliptic latitude of the source star, hence the angle between the Ecliptic plane and the plane transverse to the line of sight.
The impact parameter from Earth is therefore given by $u(t) = ||\pmb{u}_D(t) - \pmb{u}_\oplus(t)||$.

\section{Simulation of microlensing parameters}
\label{section:simulation}
To study the parallax effect in detecting microlensing events, we generated microlensing parameters from dark matter distribution models (spherical halo and thick disc). To be as general as possible, we only considered the magnifications and did not simulate measures specific to one survey, avoiding extensive experiment simulation of time sampling, photometric errors, etc. These survey characteristics are taken into account independently in an approximate way below.

We considered the two simple lens distributions that are used in many microlensing survey analyses: a galactic spherical halo model, and a dark matter thick disc.
These models correspond to two extreme lens distance distributions, from nearby to widely distributed lens distances.

\subsection{Spherical halo}
The spherical halo model consists of a spherical, isotropic, isothermal dark matter halo distribution. The mass spatial distribution is given by
\begin{equation}
\rho(r)=\rho_\odot \frac{R_c^2+R_\odot^2}{r^2+R_c^2},
\end{equation}
where $R_c = 5$ kpc is the "core" radius of the galaxy, $R_\odot = 8.5$ kpc is the distance from the Sun to the Galactic centre (GC), $\rho_\odot = 0.0079$ M$_\odot$ pc$^{-3}$ is the local dark matter density, and $r$ is the distance from the GC.
The lens velocity vector ({\bf v}) orientations are uniformly distributed, and the velocity norm probability distribution is \begin{equation}
p(v) = 4\pi v^2\left(\frac{1}{2\pi v_0^2}\right)^{3/2} e^{-\frac{v^2}{2v_0^2}}
,\end{equation}
where the velocity dispersion is $v_0=120$ km s$^{-1}$ \citep{Battaglia_2005}.
This halo model is used for its simplicity, although more complex models may better fit observations \citep{Calcino_2018} (see also Sect. \ref{section:discussion}).
\subsection{Thick disc}
We also considered a dark matter thick-disc model \citep{Moniez_2017} with mass density 
\begin{equation}
\rho_{TD}(r, z) = \frac{\Sigma}{2H}\exp{\frac{-(r-R_\odot)}{R}}\exp{\frac{-|z|}{H}}.
\end{equation}
We set the column density at $\Sigma = 35$ M$_\odot\text{ pc}^{-2}$,
the height scale at $H = 1.0$ kpc, and the radial length scale at $R = 3.5$ kpc.
The velocity of a thick-disc object can be decomposed into two components: the global rotation speed, and a peculiar velocity. 
The global disc rotation velocity depends on the distance $r$ from the GC in the cylindrical galacto-centric system,
\begin{equation}
\boldsymbol{v_\text{rot}}(r) = -v_{\text{rot,}\odot}\left[1.00767\left(\frac{r}{R_\odot}\right)^{0.0394} + 0.00712\right] \boldsymbol{u_\theta},
\end{equation}
where $v_{\text{rot,}\odot}=239$ km s$^{-1}$ is the global rotation speed at the position of the Sun \citep{Brunthaler_2011}.
The peculiar velocity distribution is described by an anisotropic Gaussian distribution characterised by the following radial, tangential, and perpendicular velocity dispersions \citep{Pasetto_2012a}:
\begin{equation}
(\sigma_r, \sigma_\theta, \sigma_z) = (56.1\pm3.8,\ 46.1\pm6.7,\ 35.1\pm3.4) \text{ km s}^{-1}
.\end{equation}
\subsection{Common parameters}
The coordinates
%\footnote{The origin is the GC, $\mathbf{X}$ points from the Sun towards the GC, $\mathbf{Y}$ points in the same direction as the velocity vector of the Sun, $\mathbf{Z}$ points towards the Galactic North Pole \citep{Gardiner_1994}\LEt{it is unusual to have discursive footnotes in a paper (urls and such are okay); please move all three of your footnotes to the main text}.}
of the global velocity vector of the LMC are \citep{Kallivayalil_2013} \begin{equation}
\boldsymbol{v}_{LMC} = (-57,\ -226,\ 221)_\textbf{XYZ} \hspace{3ex} \text{km s}^{-1},
\end{equation}
in the galacto-centric system where
the origin is the GC, $\mathbf{X}$ points from the Sun towards the GC, $\mathbf{Y}$ points in the same direction as the velocity vector of the Sun,
$\mathbf{Z}$ points towards the Galactic North Pole \citep{Gardiner_1994}.
The velocity vector of the Sun is the sum of the global Galactic rotation component and the peculiar velocity component \citep{Brunthaler_2011},
\begin{equation}
\boldsymbol{v}_\odot = (11.1, 12.24 + v_{\text{rot},\odot}, 7.25)_\textbf{XYZ} \hspace{3ex} \text{km s}^{-1}.
\end{equation}
We simulated microlensing for six masses: 0.1, 1, 10, 30, 100, and 300 $M_\odot$.
From the generated physical values we can compute the magnification curve parameters $x$, $t_E$, $\pi_E$ , and $\theta$.
$u_0$ and $t_0$ are then generated uniformly ($u_0\in\left[-2, 2\right]$, $t_0$ spanning all seasons).
We did not simulate blending (but see the discussion section).
Because we study the magnification here, we did not consider the source fluxes and assumed that fluxes are measured
with the nominal photometric precision for each survey.
We generated a few million parameter sets for lenses belonging to each dark matter structure, spherical halo and thick disc. By splitting the samples into two halves, we verified that the results from each sub-sample are compatible, which confirms that this statistics is not limiting.
Because we did not simulate light curves as obtained from a given survey, we define in the next sections the metrics by only taking the microlensing event parameters as input.

\section{Parallax effect on the event detection}
\label{section:detection}
A parallax can sometimes significantly distort the light curve of a microlensing event by making it asymmetric and/or multi-peaked with approximately yearly gaps.
Before we discuss the effect of these distortions, we briefly recall the philosophy of the historical microlensing searches. The event detection is based on a preselection that in turn is based on loose selection cuts that
request a single bump in the light curve, with little constraint on the shape.
Microlensing events are then typically identified through visual control within this sample, which is much wider than the final selection.
After this preselection, an automatic (blind) selection is applied using tighter criteria on the shape of the light curve (such as $\chi^2$ of a simple microlensing fit), in order to optimise the rejection of microlensing artefacts. At this level, genuine microlensing events identified during the preselection may be rejected because of parallax distortions.
This automatic selection is necessary to compute a detection efficiency, which is needed to measure event rate and optical depth, defined as the probability for a given line of sight to intercept a deflector's Einstein disc.

Two effects from the parallax distorsions have to be distinguished.
i) {\it Event detection:}
Here, the question is to quantify the probability of missing events with a  good signal-to-noise ratio  (S/N) during the preselection because of parallax.
ii) {\it Detection efficiency:}
The complete automatic selection is obviously more sensitive to the shape of the events, and the selected event counts used in the efficiency computing may be affected by the parallax.

To study the perturbation to the event detection, we quantified the parallax effect using a metric (distance function) between theoretical parallax and standard event light curves, and the peak counts in the light curves.
\subsection{Minimum absolute photometric difference $D_\pi$}
\label{section:distance}
Even when the light curve seen from Earth differs significantly from the light curve seen from the Sun, its shape in most cases remains almost identical to a standard light curve, but with different parameters.
%To quantify the shape distortion of the light curve, our metric is computed from the light curve of the simple microlensing event $m_\odot(t, t_E,t_0,u_0)$ to the considered parallax event $m_\oplus(t)$ by minimising the maximum absolute difference,
We quantified the shape distortion of the light curve with the parameter $D_\pi$, defined as the smallest maximum absolute difference
between the considered parallax event (seen from Earth) magnitude magnification function $m_\oplus(t)$ and a simple event (seen from the Sun) magnitude magnification function $m_\odot(t, t_E,t_0,u_0)$:
\begin{equation}
D_\pi = \min_{t_E,\ t_0,\ u_0} \{ \max_{t} |{m_\oplus(t)-m_\odot(t, t_E, t_0, u_0)}| \}.
\end{equation}

\subsection{Number of peaks}
Microlensing surveys typically include the condition of observing only one significant peak  in the prefiltering.
In addition to $D_\pi$, we therefore computed the number of distinct peaks exhibited by the parallax light curve. 
A peak is defined by an interval during which the amplification exceeds a given threshold (depending on the survey).
We estimated the number of peaks in measured light curves assuming a daily sampling rate that is better than all the past and planned cadencing. It therefore provides conservative upper numbers of peaks.
\subsection{Quantifying the probability of missing events}
We only considered events with impact parameter of the best simple microlensing curve (associated with $D_\pi$)
$|u_0|<1$, because most research algorithms require a minimum magnification of 1.34, corresponding to $u=1$ in Eq. (\ref{Amplification}).
\begin{figure}
    \centering
    \includegraphics[width=\linewidth]{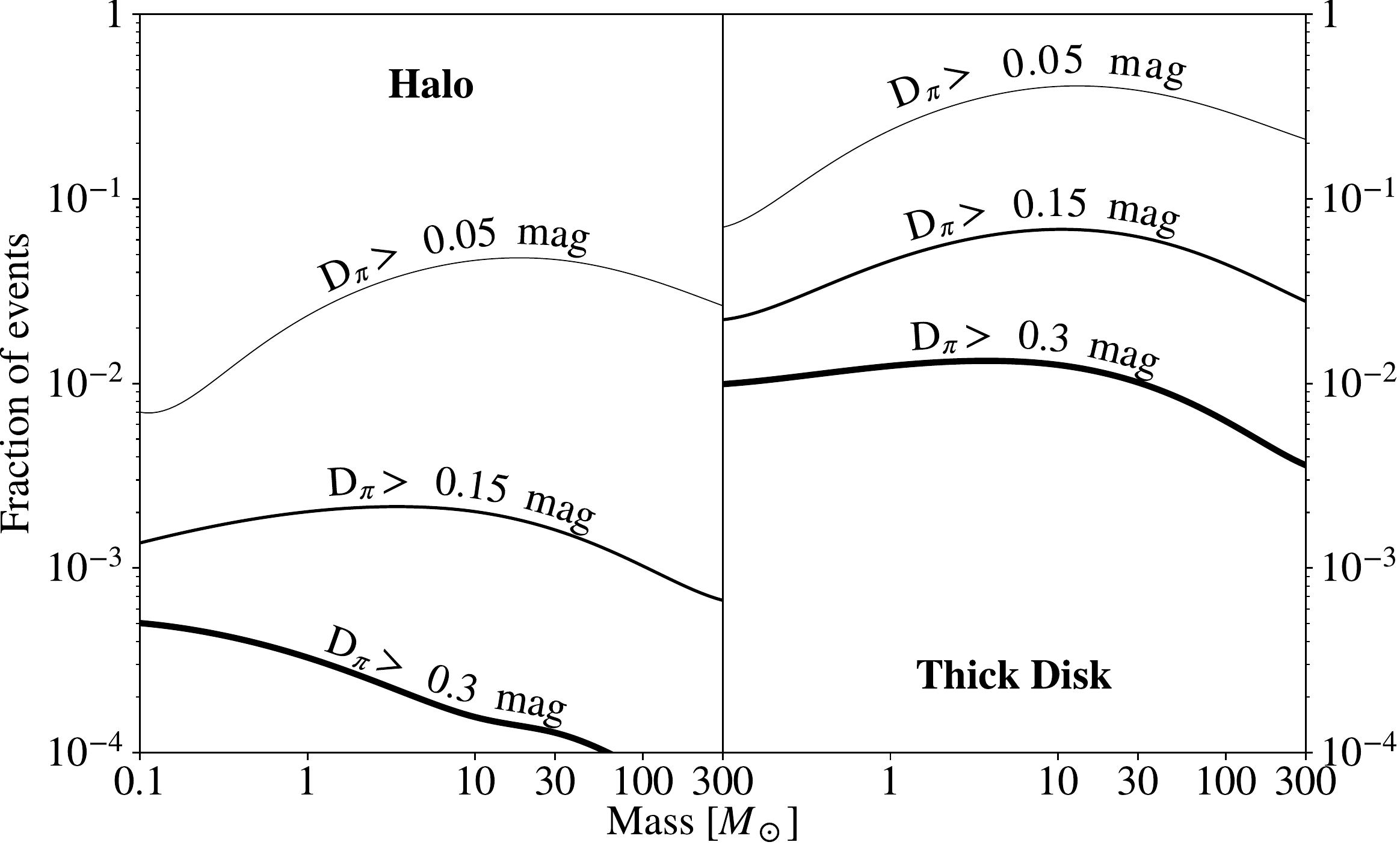}
    \caption{\it Fraction of light curves with best fit (simple event) $u_0<1$ and $D_\pi$ larger than (0.05, 0.15, 0.3) magnitude as a function of the deflectors' mass. Spherical halo dark matter model (left), and thick-disc dark matter model (right).}
    \label{fig:minmax}
\end{figure}
Figure \ref{fig:minmax} shows the fraction of events with $D_\pi$ larger than given thresholds (0.05, 0.15, 0.3 magnitudes), as a function of the deflectors' mass.
The curves corresponding to $D_\pi>0.05$ mag are the easiest to interpret because $0.05$ mag (chosen to be conservative) corresponds to the typical best photometric resolution $\sigma_\text{phot.}$ of the past surveys (EROS, MACHO, MOA, and OGLE2-3). Figure \ref{fig:minmax} shows that a maximum of $5\%$ ($40\%$) of the event light curves from the halo dark matter model (thick disc) can deviate by more than 0.05 magnitudes from a standard microlensing light curve. 
This means that at least $95\%$ ($60\%$) of the events are indistinguishable from standard events as soon as $\sigma_\text{phot.}>0.05$.
The fraction of events that deviates by more than 0.15 magnitudes is always smaller than $1\%$ for objects from the halo dark matter model, and lower than $8\%$ for the thick-disc dark matter model. 

We considered events due to lenses that belong to the thick-disc dark matter model in more detail, where deviations can be somewhat more frequent (maximum of $40\%$ with $D_\pi>0.05$ mag, maximum of $8\%$ with $D_\pi>0.15$ mag).
We estimate that the probability of finding more than one peak above any threshold between 0.05 and 0.5 magnitude during the prefiltering process is negligible (lower than $0.6\%$ for thick-disc events and even lower for halo events).
Because a common prefiltering requirement is the presence of only one significant bump in the light curve, this means that no event from lenses between 0.1 to 300 M$_\odot$ should be eliminated by this requirement, regardless of the precise shape of the bump.

The first (conservative) conclusion is that if lenses belong to a halo, the prefiltering of microlensing events is expected to never miss more than $5\%$ of the long events because of the parallax;
if lenses belong to a thick disk, the distortions may more frequently exceed $0.05$ magnitude. Nevertheless, because the usual prefiltering tolerates a shape alteration lower than $0.15$ mag and because there is no more than one peak in the light curve, this prefiltering is not expected to miss more than $8\%$ of the long events.

It appears that as soon as a criterion for the goodness of fit is used, the automatic selection algorithm efficiency has to either be loose enough to allow for $\sim 0.15$ magnitude variations (case of thick-disc lenses) or be estimated including the parallax. 
In the case of a sub-percent photometric survey such as the LSST, the required automatic detection efficiency for estimating the optical depth of the microlensing or for establishing limits on the thick-disc contribution to dark matter will need to properly account for parallax, and more specific studies are required.

\section{Potential of parallax detection}
\label{section:potential}
In the previous section, we discussed the parallax effect on the detection of microlensing events. In this section, we discuss the probability of detecting the parallax in addition to the microlensing effect. 
Such a detection offers the potential of significantly improving the constraints on the configurations of microlensing events \citep{Han_1995, Poindexter_2005, Wyrzykowski_2016}.
Moreover, a systematic search for parallax during on-going events enables triggering complementary fast-sampled, multi-colour, or any other specific observations.
\subsection{Minimum integral photometric difference}
\label{section:surveys}
To estimate the proportion of events with detectable parallax effect, we quantified the parallax significance with a pseudo-$\chi^2$ , defined as
\begin{equation}
    \widetilde{\chi}^2_\pi = N_\text{obs} + \frac{f_s}{\sigma_{phot.}^2} \min_{t_E, t_0, u_0} \int_{-\infty}^{+\infty}\left(m_\odot(t, t_E, t_0, u_0)-m_\oplus(t) \right)^2 dt
    \label{integral}
,\end{equation}
where
$m_\odot$ ($m_\oplus$) is the magnification function of the lensed source seen from the Sun (from Earth), and the
sampling frequency $f_s$, the mean photometric error $\sigma_{phot.}$, and the number of measures $N_\text{obs}=T_{obs}\times f_s$
characterise the survey.
This pseudo-$\chi^2$ is a proxy of the $\chi^2$ of the best standard microlensing fit to an hypothetically observed light curve containing $N_{obs}$ observations, sampling a true microlensing light curve $m_\oplus(t)$ (with parallax) with a constant photometric precision $\sigma_{phot.}$.

Clearly, $\widetilde{\chi}^2_\pi$ cannot account for all subtleties of a dedicated simulation, because i) this quantity does not take variations in photometric precision with magnitude or with observational conditions into account; as a consequence, this pseudo-$\chi_2$ is an averaged proxy.
ii) For simplicity, the integral is extended to infinity instead of the time interval of observation $T_{obs}$; this approximation is pessimistic because events with parallax deviations exceeding $T_{obs}$ interval will have a larger $\widetilde{\chi}^2_\pi$ than when the integral is restricted to $T_{obs}$.
iii) Time sampling is more critical for short and/or highly magnified events; we therefore trust this proxy mainly for studies of long-timescale events;
the exact time distribution of the measurements is not expected to significantly affect parallax detection for events that are several years long if there are no year-long gaps without measurements.

After minimising the integral from Eq. (\ref{integral}), we computed  $\widetilde{\chi}^2_\pi$ by setting the parameters $N_{obs}$ (or $T_{obs}$), $f_s$ , and $\sigma_{phot.}$ associated with a given survey. 
From $\widetilde{\chi}^2_\pi$,
we computed an associated significance of the improvement from standard fit to a parallax fit (in $\sigma$) and estimated the fraction of events with detectable parallax effects (at $3\sigma$ and $5\sigma$).

\subsection{Quantifying the parallax detectability}
\begin{figure}
    \centering
    \includegraphics[width=0.8\linewidth]{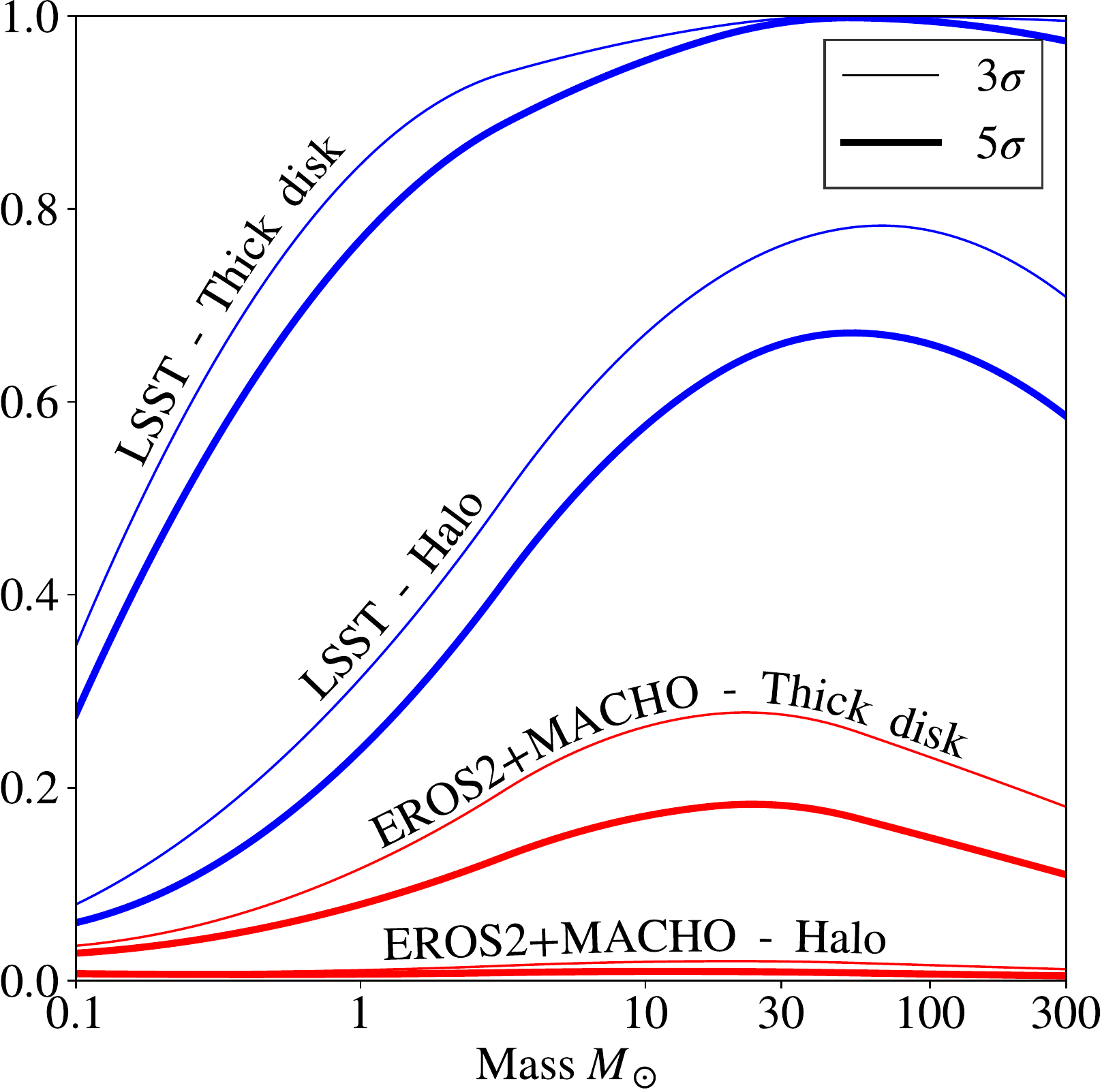}
    \caption{\it Fraction of events with parallax significance $>3\sigma$ (thin lines) and $>5\sigma$ (thick lines) as a function of the lens mass.}
    \label{fig:integral_linear_dll}
\end{figure}
Using the optimal parameters for the joint EROS2+MACHO surveys \citep{Mirhosseini_2018} ($\sigma_{phot.}=0.05$ mag, $f_s=0.2$ day$^{-1}$, $T_\text{obs}=4000$ days), and the
projected parameters for LSST \citep{LSST_design_2019, olsen2018mapping} ($\sigma_{phot.}=0.005$ mag, $f_s=0.25$ day$^{-1}$, $T_\text{obs}=4000$ days), we obtained the curves shown in Fig. \ref{fig:integral_linear_dll} for $3\sigma$ and $5\sigma$ detection. For the dark matter spherical halo model, we expect a negligible fraction of events to have a significant parallax signature in the combined database of EROS2 and MACHO;
conversely, LSST alone ({\it i.e.} with no follow-up observations) is expected to be able to detect parallaxes of more than $65 \%$ of the microlensing events in a mass range of $10$ M$_\odot$ - $300$ M$_\odot$ because the photometric precision is far better.
In a dark matter thick-disc model, the fraction of events with a measurable parallax effect is significant in both surveys (except for 0.1 M$_\odot$ deflectors for EROS2+MACHO).
The high sensitivity to parallax found for the LSST agrees well with a previous study \citep{Rahvar_2003} that assumed Hubble Space Telescope observations triggered by Earth microlensing-alert systems.
\section{Discussion}
\label{section:discussion}
We here neglected the cases of other non-standard microlensing effects (multiple sources and structured lenses), which represent a marginal statistics.
We verified the robustness of our conclusions with respect to blending:
the catalogued sources in the microlensing surveys are frequently composite objects; in this situation, the observed light curve is the superposition of a microlensing light curve and a constant one. We simulated parallaxed events with up to a $50\%$
blended contribution to the light curve and considered the same ratios as defined in Fig. \ref{fig:minmax}. We find that at least up to a $50\%$ blending level, the probability of missing events when the parallax is ignored is only marginally changed with respect to the case without blending.

Our study focused on the case of dark matter black holes belonging to a spherical halo or a thick disc. Any other reasonable model for the dark matter structure, such as  flattened halo model or a power-law model \citep{Calcino_2018}, should have a lens distance distribution and kinematical data that are a compromise between these two limiting cases. Therefore the numbers extracted from Figs. \ref{fig:minmax} and \ref{fig:integral_linear_dll} provide minimum and maximum event fractions. The conclusions relative to the prefiltering effect may thus be generalised to other halo models, and the expected fractions of detectable parallax events should be interpolated between these extreme values.
We plan to deliver the results of an equivalent study for the microlensing searches towards the Galactic plane in a forthcoming paper.
\section{Conclusions and perspectives}
Our study shows that the parallax has a negligible effect on prefiltering of long-duration (more than a few months) microlensing events towards the LMC and SMC in the case of either a dark matter spherical halo or a dark matter thick disc composed of black holes.
The parallax should be taken into account for computing the efficiency of automatic filtering algorithms, however, specifically for the next-generation experiments with their sub-percent level of photometric accuracy;
dedicated simulation based on realised observation cadences and observations is required to precisely assess this detection efficiency to extract optical depths and event rates.
For most lenses with masses higher than 10 $M_\odot$ towards the LMC and SMC, LSST-like surveys should be able to detect and quantify the parallax, allowing a better determination of the lensing configuration parameters, and a distinction between models for the dark matter structure.
\begin{acknowledgements}
We thank Olivier Perdereau and Sylvie Dagoret-Campagne for their useful comments on the manuscript. This work was supported by the Paris Ile-de-France Region.
\end{acknowledgements}
\bibliographystyle{aa}
\bibliography{citations_parallax}
\end{document}